# Single photon source characterization with a superconducting single photon detector

**Robert H. Hadfield, Martin J. Stevens, Steven S. Gruber,**

*National Institute of Standards and Technology, 325 Broadway, Boulder CO 80305, USA*
*hadfield@boulder.nist.gov*

**Aaron J. Miller,**

*Department of Physics, Albion College, Albion MI 49924, USA*

**Robert E. Schwall, Richard P. Mirin, Sae Woo Nam**

*National Institute of Standards and Technology, 325 Broadway, Boulder CO 80305, USA*

**Abstract:** Superconducting single photon detectors (SSPD) based on nanopatterned niobium nitride wires offer single photon counting at fast rates, low jitter, and low dark counts, from visible wavelengths well into the infrared. We demonstrate the first use of an SSPD, packaged in a commercial cryocooler, for single photon source characterization. The source is an optically pumped, microcavity-coupled InGaAs quantum dot, emitting single photons on demand at 902 nm. The SSPD replaces the second silicon Avalanche Photodiode (APD) in a Hanbury-Brown Twiss interferometer measurement of the source second-order correlation function, $g^{(2)}(\tau)$. The detection efficiency of the superconducting detector system is >2 % (coupling losses included). The SSPD system electronics jitter is 170 ps, versus 550 ps for the APD unit, allowing the source spontaneous emission lifetime to be measured with improved resolution.

## 1. Introduction

Single photon sources and detectors are key enabling technologies in the growing field of quantum information [1]. The ideal detector for these applications would have high speed, zero dark counts, and high quantum efficiency at the wavelength of interest. Silicon Avalanche Photodiodes (Si APDs) are the detector of choice for single photon counting at visible wavelengths [2,3,4], with high detection efficiency (up to 76 % at 700 nm) and low dark counts (~100 Hz) [5]. The lowest timing jitter reported is 20 ps full width at half maximum (FWHM) [6], but typical values for commercial units are 10 to 20 times higher. Count rates up to 5 MHz are commonly achieved. For near-infrared wavelengths such as the standard telecommunication wavelengths (1310 nm and 1550 nm) single photon counting is achieved through use of InGaAs APDs [7]. These require cooling to 200 K, offer reduced detection efficiency (20-30 %), and are limited to count rates of ~100 kHz. Furthermore, gating is essential to reduce the high dark count rate. Typical InGaAs APD units offer sub-nanosecond jitter [8]. Jitter can be reduced (down to 56 ps) at the cost of elevated dark counts [9,10].

Two new classes of superconducting detectors with single photon counting capability have recently emerged, offering considerable advantages over conventional semiconductor detector technologies. The single photon counting capability of both these superconducting detector technologies extends well into the infrared. Transition edge sensors (TES) [11,12] offer very high detection efficiency (up to 89 %) and zero dark counts, but are slow (much slower than standard fiber telecommunication clock rates) and operate at 100 mK. Superconducting single photon detectors (SSPD) [13-16] have lower detection efficiency (reportedly up to 20% at visible wavelengths) [14] and finite dark counts, but are potentially extremely fast (approaching telecommunication clock rates ~1 GHz) [13]. The best reported jitter is 20 ps FWHM [15]. Since SSPDs can be operated at temperatures near 4 K, our goal is to package the SSPD devices in a practical, turnkey, cryogen-free system that is assembled from commercially available components, and can be used in quantum information and quantum optics experiments.

## 2. Superconducting single photon detector system

The SSPD device itself is a narrow superconducting wire embedded in a 50 Ω transmission line. The superconducting track is current biased just below its critical current $I_C$. When a photon strikes the wire, a hot spot is momentarily formed. A voltage is developed briefly across this resistive section of the track, causing a high-speed voltage pulse to propagate along the transmission line. The early devices, consisting of a short section of track, suffered from low detection efficiency [13], owing to the difficulty of coupling light to such a small detector area. The SSPDs of the most recent generation boost detection efficiency through a 100 nm width meander line with 200 nm pitch, covering a 10 μm x 10 μm area [14]. In the system that we report here, a low-noise current source is used to bias the detector, and commercial room temperature rf amplifiers with adequate bandwidth and sufficiently low noise figure amplify the pulses generated by the detector. These voltage pulses can be observed on an oscilloscope, read out on digital counter, or converted into a digital logic pulse.

Light is coupled to the detector by a method previously developed for TES [11,12]. Each detector is mounted on a brass block chip carrier. Aluminum wirebonds connect the on-chip coplanar waveguide to an SMA connector. High-speed coaxial cables that are heat sunk at 2.9 K and 40 K conduct the signal from the detector to room temperature. A polished fiber, mounted in a ceramic ferrule, is fixed in a second brass block positioned over the chip carrier. The fiber in the ferrule holder is aligned over the detector by viewing the transmission of fiber-coupled 1550 nm light through the chip with an infrared microscope. The position of the fiber/fiber holder is adjusted such that the detector blocks the light. This method allows alignment of the fiber to the chip with a precision of a few micrometers and maintains that alignment over repeated thermal cycles. Single mode optical fiber enters the cryostat via an epoxy feed through.

The SSPD will operate effectively only at a constant temperature. The detection efficiency and dark count rate are dependent on both the operating temperature and the bias current as a fraction of device critical current [13-16]. The critical current of the superconducting track is, in turn, a sensitive function of temperature [17] – therefore temperature variations will result in unstable detection efficiency and dark count rate at fixed bias current. For the SSPD system we have elected to use a compact closed cycle cryogen free Gifford-McMahon type cryocooler [18]. This refrigerator provides 0.1 W of cooling power at 4 K and 5 W at 60 K, using a 1.5 kW air-cooled compressor. In our system the heat loads are minimal, and the first and second stages of the cooler operate at ~40 K and 2.9 K, respectively. The refrigerator has sufficient capacity to cool a number of detectors simultaneously. However, the temperature at the coldest stage fluctuates +/- 0.3 K peak-to-peak at 1.2 Hz. In the system that we report here, passive temperature stabilization of the detector assembly is achieved by weakly coupling the detector stage to the cold stage. Using a small volume of lead to provide heat capacity on the detector stage and a thin copper wire to provide a controlled thermal conductance between detector stage and cold stage, we were able to isolate the detector from the cold stage with a thermal time constant, $C/g$ of ~100 sec. This arrangement reduced temperature variations at the detector to ~0.02 K peak-to-peak.

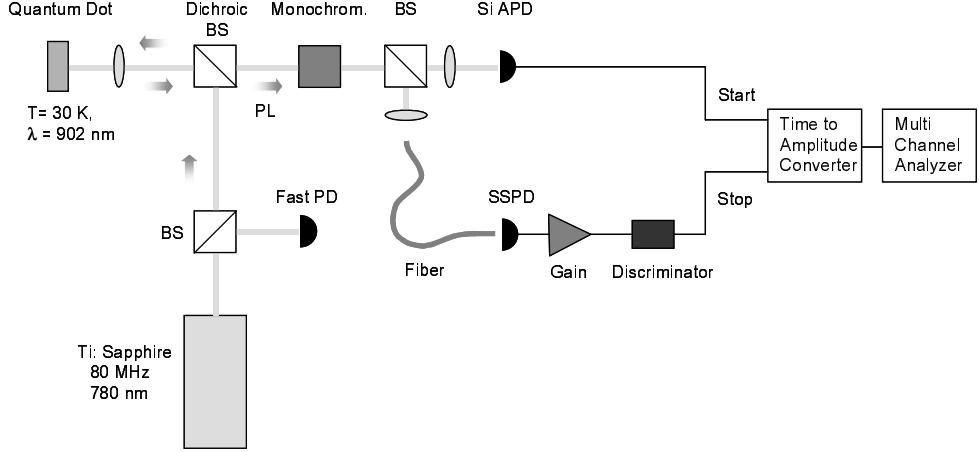

Fig. 1. Schematic of the coincidence measurement setup used for single photon source characterization. The Si Avalanche Photodiode (APD) is used as the start trigger for the time-to-amplitude converter/multichannel analyzer and the Superconducting Single Photon Detector (SSPD) is used as the stop. In the carrier lifetime measurement either the SSPD or APD is used as the start, and the stop trigger comes from the fast photodiode (PD) monitoring the laser output.

Table 1. Detector performances in this experiment

| Detector | Counts per second | Dark counts per second | Detection Efficiency $\lambda$=902 nm | Jitter (ps) |
| --- | --- | --- | --- | --- |
| Si APD | 13,000 | 100 | 38%* | 550 |
| SSPD | 600 | 50-100 | 2%† | 170‡ |

*Manufacturer specification    †Coupling losses included    ‡Limited by electronics

## 3. Coincidence measurement

A schematic of the experimental setup is shown in Fig. 1. The single photon source is an individual InGaAs quantum dot placed inside a micropillar cavity [19,20]. The optical cavity is formed by a pair of GaAs/AlAs distributed Bragg reflectors grown above and below the dot. The cylindrical micropillar, which is ~6 µm tall and ~2 µm in diameter, is defined with photolithography and a reactive ion etch. The dot is optically pumped with a Ti:Sapphire laser that produces ~1 ps pulses with a center wavelength $\lambda$ = 780 nm, at 82 MHz repetition rate. The sample is cooled to ~30 K in a liquid He flow cryostat. At this temperature, the dot emits single photons on demand at $\lambda$ = 902 nm, on resonance with the micropillar cavity mode (see Fig. 2). The light emitted from the micropillar is collected with an objective lens, spectrally filtered with a grating monochrometer and directed into a Hanbury-Brown Twiss (HBT) interferometer [21]. The HBT interferometer consists of a beamsplitter (BS) and two detectors: one output of the BS is focused onto a Si APD, and the other is coupled to a single mode fiber whose output is sent to the SSPD. Detector count rates and performances are shown in Table 1. The detection efficiency (DE) of the SSPD system is deduced from the count rate relative to the Si APD (DE 38 %). The range over which SSPD performance can be tuned is discussed in Section 5. Fiber coupling losses [22] are included in the SSPD system DE – thus the 2 % value in the table represents the lower limit of the DE of the detector itself at $\lambda$ = 902 nm. As the coupling efficiency into the SSPD system may be as low as 20 %, the true DE of the SSPD at 902 nm may be up to 10 %.

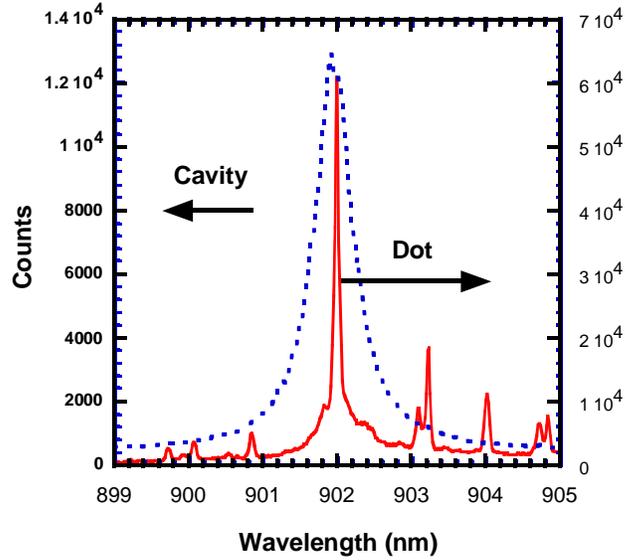

Fig. 2. Emission from cavity embedded quantum dot at 30 K. At high pump power (dotted blue trace) the cavity resonance at ~ 902 nm dominates. At low pump power (solid red trace) the main photoluminescence peak is aligned to the cavity resonance

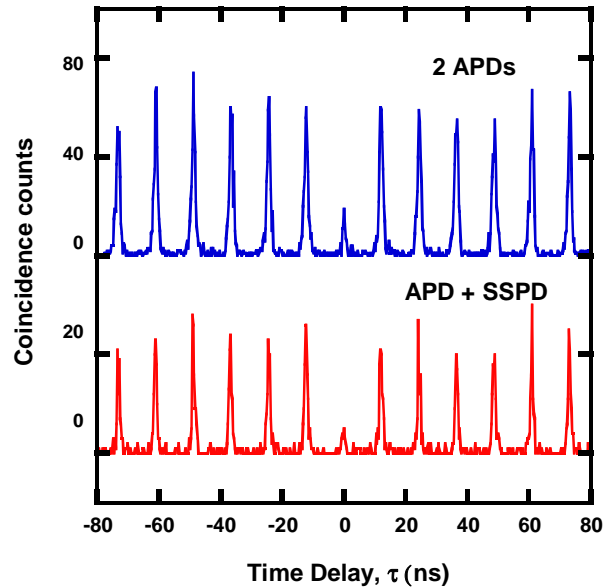

Fig. 3. Coincidence counts versus delay time histogram for an InGaAs single photon source at $\lambda = 902$ nm measured with a double APD (upper, blue trace) and an SSPD/APD (lower, red trace) configuration. The coincidences are proportional to the second order correlation function of the source, $g^2(\tau)$. In both cases the zero delay peak area is suppressed to 24 % of the area of the other peaks (averaged over a wider range than shown), indicating a four-fold decrease in multiphoton emission relative to a Poissonian source of equal intensity. The non-zero delay peaks in the SSPD/APD trace are 17 % narrower than those of the double APD trace, owing to the lower jitter of the SSPD relative to an APD.

Voltage pulses from the SSPD are amplified by an rf amplifier chain and converted to logic pulses by means of a leading-edge discriminator that is typically used for photomultiplier tubes. The APD unit similarly produces logic pulses. Pulses from the two detector units are fed into a digital time-to-amplitude converter (TAC). The Si APD starts a timer and the SSPD stops it. A multichannel analyzer (MCA) is then used to produce a count versus time interval ($\tau$) histogram (Fig. 3), which is proportional to the second order correlation function of the source, $g^{(2)}(\tau)$. The area of the peak at $\tau = 0$ divided by the average area of the surrounding peaks is equal to the second order correlation function of the source at zero delay, $g^{(2)}(0)$. For the quantum dot emission at 902 nm we find $g^{(2)}(0) = 0.24 \pm 0.03$. This indicates a four-fold reduction in the probability of generating more than one photon in a pulse, relative to a classical, Poissonian source of the same intensity. We have repeated this measurement using a Si APD in each arm of the HBT interferometer, and found $g^{(2)}(0) = 0.24 \pm 0.05$. The width of the peaks in the histogram is determined by the detector instrument response function (IRF) convolved with the spontaneous emission lifetime (see next section). The peaks in the SSPD/APD configuration were 17 % narrower, than in the double APD configuration, owing to the lower jitter of the SSPD. If a second SSPD channel were available, we should expect a 40 % narrowing of the peaks.

## 4. Spontaneous emission lifetime measurement

Another advantage of the SSPD over the APD is demonstrated by measuring the spontaneous emission lifetime of the quantum dot by a standard time-correlated single photon counting technique in reverse start-stop mode. The experimental setup in Fig. 1 was modified so that either the SSPD or the Si APD provided the start signal, while the 82 MHz clock signal from the fast photodiode monitoring the pump beam provided the stop. Results are shown in Fig. 4. The sharp initial rise in the SSPD measurement, in contrast to the much slower rise in the APD data, is due to the improved time resolution available from the SSPD.

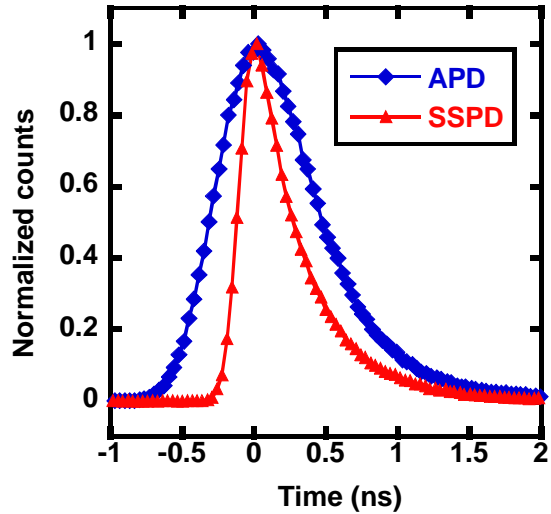

Fig. 4. Time-resolved measurements of spontaneous emission lifetime at 902 nm on resonance. The red triangles are data acquired with an SSPD, the blue diamonds with an APD. The plots are fitted (solid lines) with an exponential decay (time constant 0.37 ns) convolved with a Gaussian approximation to the instrument response function (IRF). The SSPD IRF has a FWHM of 170ps – this is wider than SSPD jitter measurements reported elsewhere and may be broadened by our rf amplifiers and TCA/MCA electronics. The APD IRF is nevertheless significantly wider (FWHM 550 ps).

The data in Fig. 4 are fitted (solid lines) with an exponential decay (time constant 370 ps) convolved with a Gaussian approximation to the instrument response function (IRF) of each detector (the IRF can be measured by illuminating the single photon detector with the Ti:Sapphire laser and stopping on the fast photodiode). The SSPD system has a FWHM of 170 ps; this IRF is wider than reported elsewhere [15] and is probably broadened by our rf amplifier chain and TCA/MCA electronics. Nevertheless the SSPD system has a significantly narrower IRF than the Si APD unit (the minimum jitter for we have observed for this APD unit under optimal alignment is ~350 ps – in this experiment the measured jitter was 550 ps).

## 5. Detection efficiency and dark counts at λ=902 nm and 1550 nm

As mentioned in Section 3 the SSPD system detection efficiency (DE) at 902 nm can be determined relative to the known performance of the Si APD unit. In the spontaneous emission lifetime measurement the signal to noise (SNR) achieved using the SSPD can be tuned by altering the bias current. The SSPD DE at 902 nm and dark count rate are shown in Fig. 5. Maximum detection efficiency (up to 3 %) and dark count rate (1 kHz) is obtained when $I_{Bias}$ approaches $I_C$. Below $I_C$ the dark count falls more rapidly than the detection efficiency, allowing us to achieve a 2 % detection efficiency at ~100 Hz dark count rate in this experiment.

Determining the DE of the fiber-coupled detector is straightforward at standard telecommunications wavelengths. A gain-switched 1550 nm laser (drive frequency $f = 100$ kHz) is coupled to the device via a programmable digital attenuator. A relatively low drive frequency is chosen to allow the effect of the detector dead time (~10 ns) to be neglected. A conventional power meter is used to measure the average power output from the pulsed laser. Using the known drive frequency and wavelength, we can deduce $\mu$, the mean number of photons per pulse. At a given current bias $I_{Bias}$ the number of counts per second $R$ is recorded as $\mu$ is attenuated. The single photon DE, $\eta$ coupled to a source with a Poissonian photon number distribution can be extracted by fitting to the relation $R=D + f(1-\exp(-\eta\mu))$, where $D$ is the dark count rate. Figure 5 shows the system DE, $\eta$ versus dark count rate $D$ at both 902 nm and 1550 nm. Both $\eta$ and $D$ are maximized as $I_C$ is approached. Intrinsic detection efficiencies of up to 10 % for the same detectors at 1550 nm have been reported elsewhere [15]. For our packaged fiber coupled detector the maximum observed DE is ~ 1 %. At shorter wavelengths we see an improvement in $\eta$ owing to the higher photon energy. The SSPD is current-biased close to the critical current of the superconducting track and is thus susceptible to dark counts arising from electronic and thermal fluctuations even in a light-tight environment. Gating the discriminator response window in time can minimize the effect of the intrinsic dark count rate [23].

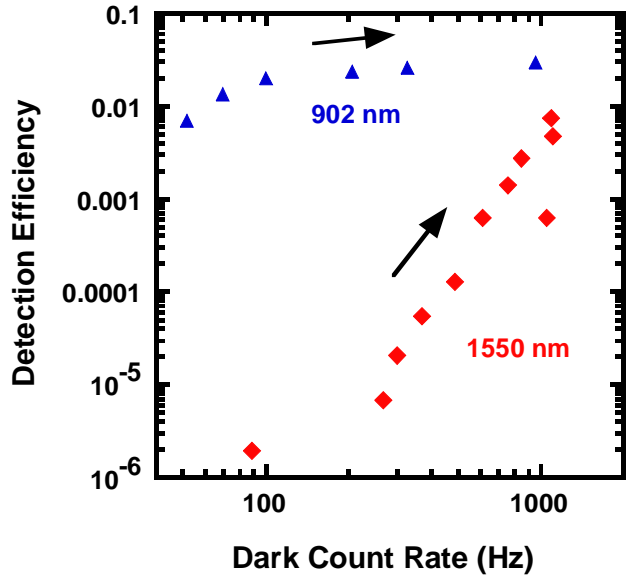

Fig. 5. SSPD system detection efficiency versus dark counts at 902 nm (blue triangles) and 1550 nm (red diamonds). The arrows indicate the direction of increasing bias towards $I_C$. The data covers the approximate range in $I_{bias}$ from 60% to 95% $I_C$.

## 5. Conclusion

We have successfully employed a superconducting single photon detector system for the first time in the characterization of a single photon source at 902 nm. The performance currently achieved in our prototype SSPD system falls short of the best reported [15, 16]; nevertheless the reduced jitter of the SSPD channel allows us to make improved measurements of source $g^{(2)}(\tau)$ and carrier lifetime. Single photon sources in the near infrared are already a reality [24-26]. Unlike Si APDs, this detector retains single photon counting capability into the infrared, making it a promising candidate for single photon counting applications at conventional telecommunications wavelengths, such as long distance fiber-based quantum key distribution. We anticipate improved performance in the future for multiple SSPD channels packaged in a single cryocooler.

**Acknowledgements**
We gratefully acknowledge support from the United States Department of Commerce, DARPA and BBN technologies. We thank Professors Gregory Gol'tsman and Professor Roman Sobolewski for providing the original detectors used in this experiment and for useful technical discussions. We also thank Norm Bergren for technical assistance.